\documentclass[twocolumn,prl,showpacs,floatfix]{revtex4}
\usepackage{graphicx}
\usepackage{bm}

\begin{document}

\title{\textbf{Direct Observation of Charge Order in Triangular
Metallic AgNiO$_2$ \\by Single-crystal Resonant X-ray Scattering}}
\author{G.L.\ Pascut$^{1,2}$, R. Coldea$^{1}$, P.G. Radaelli$^{1}$,
A. Bombardi$^{3}$, G. Beutier$^{3}$, I.I. Mazin$^{4}$, M.D.
Johannes$^{4}$, and M. Jansen$^{5}$} \affiliation{$^1$Clarendon
Laboratory, University of Oxford, Parks Road, Oxford OX1 3PU,
United Kingdom\\ $^2$H.H. Wills Physics Laboratory, University of
Bristol, Tyndall Avenue, Bristol, BS8 1TL, United Kingdom\\
$^3$Diamond Light Source Ltd., Harwell Science and Innovation
Campus, Didcot, Oxfordshire, OX11 0DE, United Kingdom\\
$^4$Code 6393, Naval Research Laboratory, Washington, D.C. 20375\\
$^5$Max-Planck Institut f\"{u}r Festk\"{o}rperforshung,
Heisenbergstrasse 1, D-70569 Stuttgart, Germany}

\date{\today }

\begin{abstract}
We report resonant X-ray scattering measurements on a single
crystal  of the orbitally-degenerate triangular metallic
antiferromagnet $2H$-AgNiO$_2$ to probe the spontaneous transition
to a triple-cell superstructure at temperatures below $T_{\rm
S}=365$~K. We observe a strong resonant enhancement of the
supercell reflections through the Ni K-edge. The empirically
extracted K-edge shift between the crystallographically-distinct
Ni sites of 2.5(3) eV is much larger than the value expected from
the shift in final states, and implies a core-level shift of
$\sim$ 1 eV, thus providing direct evidence for the onset of
spontaneous honeycomb charge order in the triangular Ni layers. We
also provide band-structure calculations that explain
quantitatively the observed edge shifts in terms of changes in the
Ni electronic energy levels due to charge order and hybridization
with the surrounding oxygens.
\end{abstract}

\maketitle

\begin{figure}[tbhp]
\begin{center}
\includegraphics[width=7cm,bbllx=79,bblly=625 ,bburx=520,
bbury=810,angle=0,clip=]{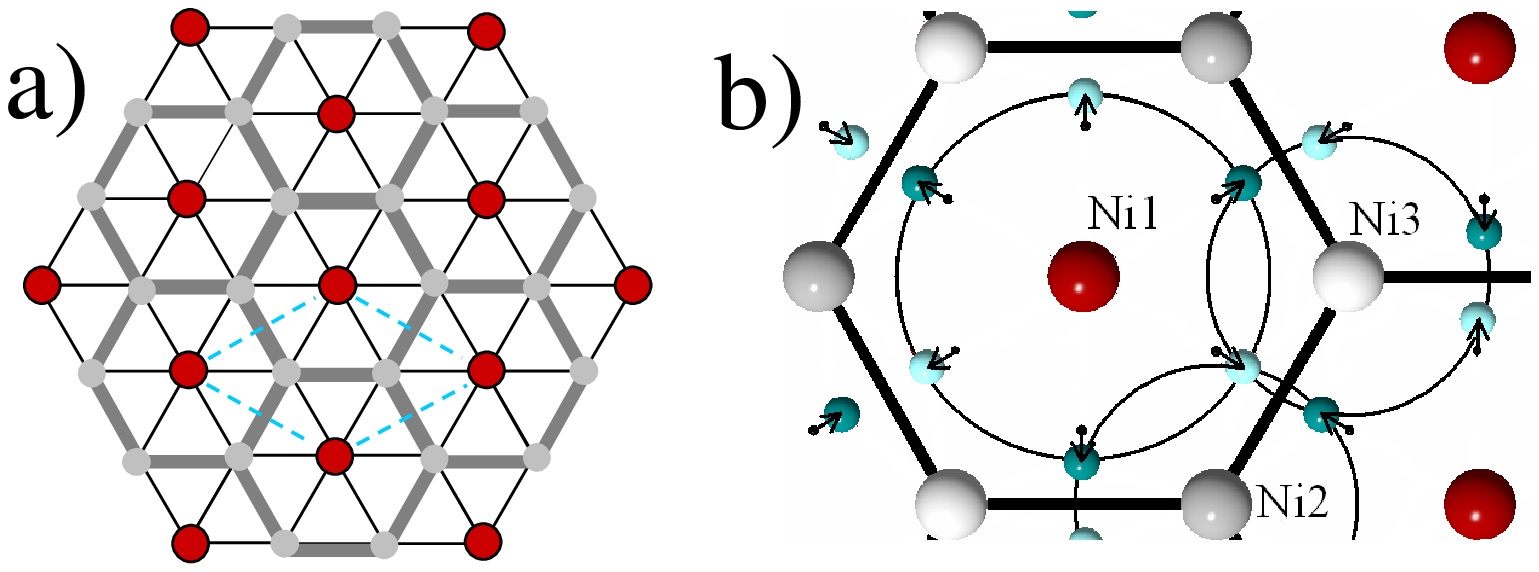}
\includegraphics[width=3.93cm,bbllx=67,bblly=357 ,bburx=348,
bbury=564,angle=0,clip=]{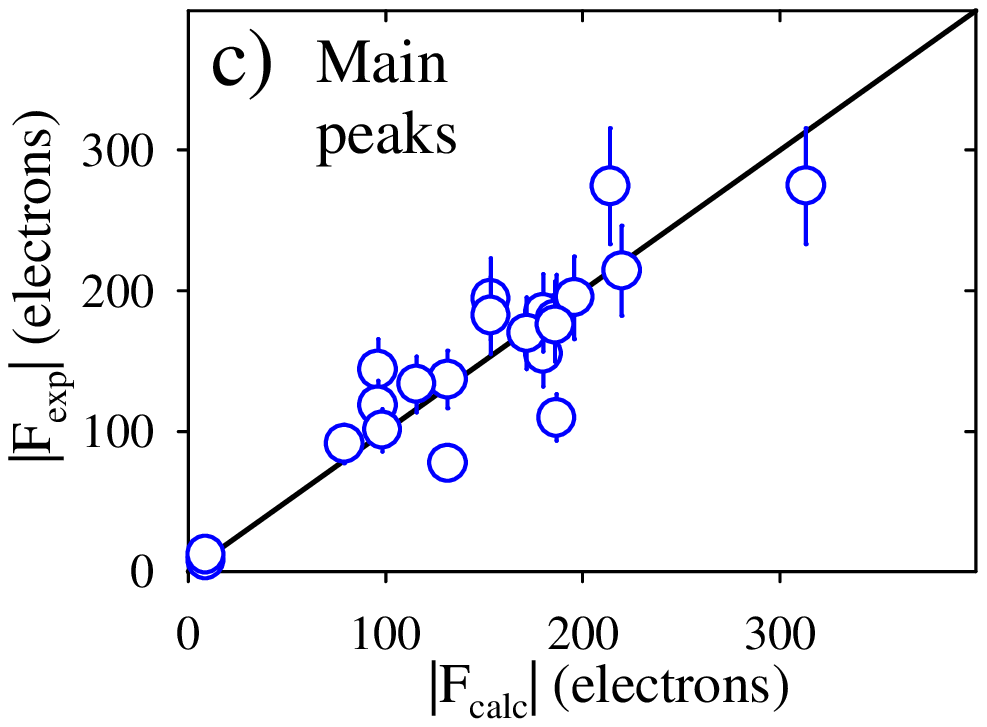}\hspace{0.25cm}
\includegraphics[width=3.85cm,bbllx=114,bblly=357 ,bburx=388,
bbury=564,angle=0,clip=]{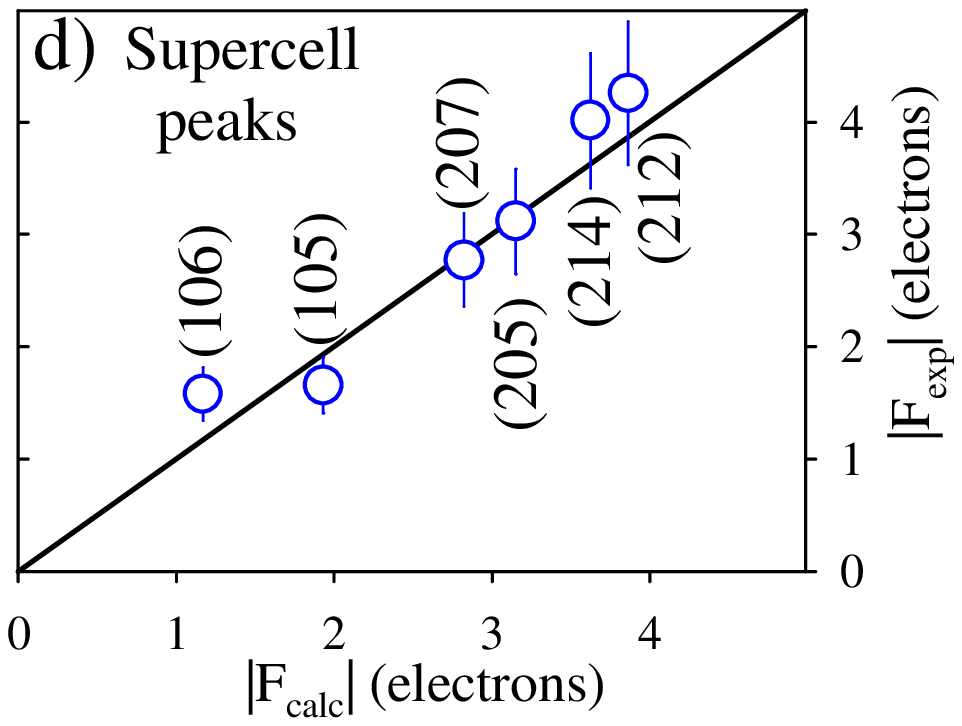} \caption{\label{fig1} (color
online) a) Schematic of the honeycomb CO in the triangular
lattice: red (dark) balls are electron-rich Ni1 sites nested
inside a honeycomb (thick contour) of electron-depleted Ni2,3
sites (gray balls).  The dashed rhombus is the CO supercell. b)
Oxygen breathing mode around electron-rich and -depleted Ni sites.
c) and d) Observed vs. calculated off-resonance (8.3 keV) $|F|$
for main and supercell peaks.}
\end{center}
\end{figure}

Since the discovery of the Verwey transition in 1939
\cite{magnetite_nature}, electronic ordering in oxides has been
extensively studied, as more evidence progressively emerged of its
importance in underpinning phenomena such as metal-insulator
transitions, colossal magnetoresistance and possibly
high-temperature superconductivity. Orbital ordering (OO) in
particular, whereby orbital degeneracy is lifted either by a
spontaneous lattice distortion driven by the (cooperative)
Jahn-Teller (JT) effect \cite{JT}, or by similar orbital physics
\cite{KK}, has become a cornerstone of our understanding of
oxides. This was hitherto considered ubiquitous in both band and
Mott insulators containing ``JT active'' ions. For this reason,
the recent proposal \cite{ewa} of a radically different type of
electronic ordering in the weakly metallic $2H$-AgNiO$_2$ came
initially as a surprise. In this scenario, orbital degeneracy at
JT-active low-spin Ni$^{3+}$ would be lifted through \emph{charge
disproportionation and charge ordering} (CO) rather than OO, in
sharp contrast with the closely related (but insulating) nickelate
NaNiO$_2$ \cite{Chappel}, a rather conventional JT system.
Although CO was observed before, e.g. at the simultaneous
metal-insulator and magnetic ordering transition in rare-earth
perovskite nickelates \cite{Alonso,Staub}, elucidating the physics
of $2H$-AgNiO$_2$ is crucial in establishing a pure CO (decoupled
from magnetic ordering \cite{Balents}) as an alternative paradigm
to the Jahn-Teller mechanism in ``weakly'' metallic systems close
to the Mott transition \cite{Mazin,ewa}, which \emph{remain
metallic} in the CO phase. Although the original clue for CO was
from the magnetic structure below $T_{\rm N}=19.7$\ K, the primary
indication was \emph{structural}, and relied on an accurate
neutron refinement of an oxygen breathing mode around the
different Ni sites (Fig.\ 1b) in the $\sqrt{3} a_0 \times \sqrt{3}
a_0 \times c$ supercell structure at $T < T_{\rm S}=365$\ K. The
pattern of charges (``honeycomb'' CO - see Fig.\ 1a) and the
amount of charge (formal valence) on each site can be estimated by
analysing the Ni-O bond lengths through the so-called Bond Valence
Sum (BVS) method \cite{Brown}, which yields Ni1$^{2.42+}$ and
Ni2,3$^{3.07+}$, see Fig.\ \ref{fig1}b), values that are fairly
typical of amounts of charge disproportionation in strongly
charge-ordered systems. The CO scenario was also supported by band
structure calculations in the Local Density Approximation (LDA)
\cite{ewa}. Nonetheless, obtaining direct experimental evidence of
CO \emph{independently} of the pure structural signature is of
crucial importance in strengthening the foundations of this new
oxide phenomenology.

Here we present resonant Ni-K-edge x-ray scattering data on a
single crystal of $2H$-AgNiO$_2$ to probe directly the electronic
order at the Ni sites.  This technique has been widely employed in
the past to explore charge and orbital order \cite{Murakami}
because of its sensitivity to subtle differences in the electronic
environments.  In particular, Ni K-edge resonant scattering probes
primarily dipole-allowed transitions from the core $1s$ levels,
which shift in response to the amount of charge on the ion, to the
empty $4p$ band, which is strongly sensitive to changes in the
coordination environment.  From our data we extract in an unbiased
way the anomalous scattering factors of the different Ni sites. By
comparing the experimental results with band structure
calculations, we show that the $4p$ level shift accounts for just
over a half of the edge shift, implying a core-level shift of
$\sim 1$ eV that provides direct evidence of honeycomb CO.  These
results are also quantitatively consistent with the amount of
charge disproportionation predicted by LDA calculations.

\begin{figure}[tbhp]
\begin{center}
\includegraphics[width=7.5cm,bbllx=100,bblly=150 ,bburx=478,
bbury=738,angle=0,clip=]{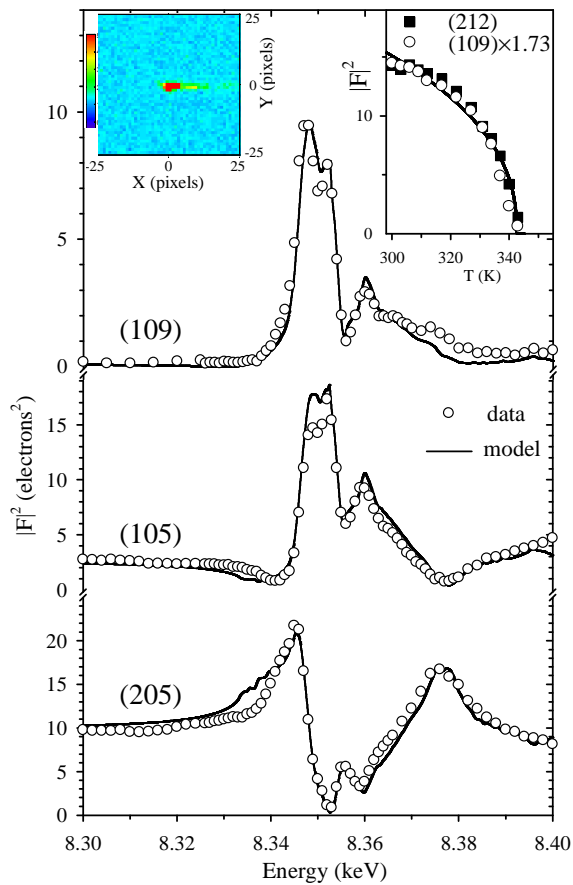} \caption{\label{fig_spectra}
(color online) Energy-dependent intensity of three supercell
reflections at 300~K near the Ni K-edge. (solid lines are model
calculations eq.~(\ref{eq_sf})). Top-left inset: example of raw 2D
pixelated area detector data for (109) near resonance. The
horizontal (vertical) angular coverage is $\approx 1^{\circ}
(0.7^{\circ})$. Top-right inset: Temperature-dependence of peak
intensities near resonance (8.347~keV) for (109), dominated by Ni
charge order (open symbols, scaled) and for (212), with
contributions from both Ni CO and Oxygen displacements (solid
squares). Solid line is a guide to the eye. The transition
temperature is offset with respect to the specific heat and
neutron diffraction value \cite{ewa} ( 343(1)~K vs. 365(5)~K),
probably due to X-ray beam heating effects.}
\end{center}
\end{figure}

Our sample was a small single-crystal platelet of $2H$-AgNiO$_2$
of diameter $\sim 70\mu$m in the $ab$ plane and thickness
$~\sim$20 $\mu$m, extracted from a polycrystalline
batch\cite{Soergel05}. A full single-crystal X-ray diffraction
pattern collected at 300 K using an Oxford Diffraction Mo-source
diffractometer confirmed the hexagonal unit cell found previously
(space group P$6_322$ with $a=b=\sqrt{3}a_0=5.0908(1)$~\AA,
$c=12.2498(1)$~\AA) where supercell peaks are indexed by $(h,k,l)$
with $h-k=3n+1$ or $3n+2$ and correspond to the oxygen breathing
order pattern shown in Fig.~\ref{fig1}b). Synchrotron resonant
X-ray diffraction (RXD) was performed at the Diamond Light Source
in Didcot, UK on beamline I16 operated with a Si (111)
double-crystal monochromator ($\Delta E/E \approx 10^{-4}$ at 8.35
keV). The azimuth of each supercell reflection was chosen such
that the incident polarisation lay as close as possible to the
$ab$ plane of the crystal (see below). The sample temperature was
controlled via a nitrogen gas flow. Intensities were collected by
a 2D Pilatus detector and a typical intensity map for a supercell
Bragg reflection is shown in Fig.~\ref{fig_spectra} top-left
inset. Total peak intensities were extracted by integrating the
counts over rocking-curve scans around the peak position. After
all corrections \cite{correction}, the off-resonance structure
factors squared $|F|^2$ at 300 K for all measured main peaks, as
well as supercell peaks agreed very well with the calculated
values, see Fig.~\ref{fig1}c-d), enabling us to convert scattering
intensities into $|F|^2$ in absolute units.

Fig.~\ref{fig_spectra} shows the energy-dependent $|F|^2$ for
three supercell reflections (out of the 6 measured), which are
representative of the different types of resonant responses. In
particular the (109) peak is entirely due to anomalous scattering
from Ni as it is practically absent off-resonance. The middle and
bottom panels show the (105) and (205) supercell peaks, which have
a sizeable oxygen contribution, as evidenced by the the off-edge
scattering.

In the dipolar approximation the atomic scattering factor near the
K-edge absorbtion energy $E_A$ is
\begin{equation}
f(Q,E)=f^0(Q)+f'(E) +if''(E), \label{eq_f}
\end{equation}
where $f^0$ is the conventional (Thomson) $Q$-dependent scattering
factor \cite{tables}, and $f'$, $f''$ are the real and imaginary
parts of the anomalous scattering factor, respectively. In
essence, the rich structures observed in Fig.~\ref{fig_spectra}
are due to the fact that $E_A$ of the electron-rich ${\rm Ni}1$ is
slightly different from that of electron-depleted ${\rm Ni}2,3$
providing a strong energy-dependent contrast between the two
sites. More specifically, the $|F|^2$ for the (109), (105) and
(205) peaks can be written as
\begin{equation}
|F(\bm{Q},E)|^2=3\left[\mp A_{\bm{Q}}f^0_{\rm{O}}/\sqrt{3} +
\Delta f_{\rm Ni}^0 +\Delta f_{\rm Ni}'\right]^2 +3\Delta
f_{\rm{Ni}}''^2, \label{eq_sf}
\end{equation}
where the upper sign applies for (109) and (105), and lower sign
for (205). The dominant term off-resonance is
$A_{\bm{Q}}f^0_{\rm{O}}$,  the energy-independent structure factor
for the oxygen breathing displacement order, which is weak for
supercell peaks like (109) and strong for (205) and (105). In
contrast, the terms $\Delta
f_{\rm{Ni}}'=f_{\rm{Ni1}}'(E)-f_{\rm{Ni3}}'(E)$ and $\Delta
f_{\rm{Ni}}''=f_{\rm{Ni1}}''(E)-f_{\rm{Ni3}}''(E)$ have the same
value for different reflections but are strongly energy-dependent
and are due to the contrast between the anomalous atomic factors
for Ni1 inside the honeycomb and Ni3 on the honeycomb
(contributions from Ni2 cancel by symmetry). The remaining term
$\Delta f_{\rm Ni}^0=f_{\rm{Ni1}}^0-f_{\rm{Ni3}}^0$ is the Thomson
charge contrast, and can essentially be neglected since the
difference in electronic density is small and highly delocalised.
All contributions to eq.~(\ref{eq_sf}) cancel in the
high-temperature phase above $T_{\rm S}$, since all Ni sites
become equivalent and the oxygens are not displaced from their
ideal positions. This can be shown by plotting the intensity of
different superlattice reflections 
as a function of temperature (top-right inset of
Fig.~\ref{fig_spectra}). These data also show that the order
parameter obtained from the purely-resonant peak (109) (``CO order
parameter'') is essentially the same as extracted from the
intensity of the (212) supercell peak, which has comparable
contributions from both Ni CO and Oxygen displacements, indicating
that, as expected, CO and displacements are strongly coupled and
both go to zero together at $T_{\rm S}$.

It is important to point out that, because of the asphericity of
the electron density, $f'$ and $f''$ in eq.~(\ref{eq_f}) are in
general not scalar quantities, but depend on the orientation of
the incident and scattered polarisations with respect to the
crystallographic axes, as expressed through the general tensor
relation $f(E)=\sum_{\alpha,\beta}\hat{f}^{\alpha\beta}
\epsilon_{\alpha}\epsilon'_{\beta}$ (valid for both $f'$ and
$f''$), where $\bm{\epsilon}$ and $\bm{\epsilon'}$ are the
incident and scattered polarisation vectors.  In $2H$-AgNiO$_2$
the local three-fold symmetry axis at \emph{each} Ni site
restricts $\hat{f}'$ and $\hat{f}''$  to have a diagonal form in
the orthogonal reference frame $xyz$ where $x
\parallel a$ and $z \parallel c$, such that
$\hat{f}=\left[f_{\perp} ~~ 0 ~~ 0; 0 ~~ f_{\perp} ~~ 0; 0 ~~ 0 ~~
f_{\parallel}\right]$, where ${\parallel}$ is along the $c$ axis
and ${\perp}$ is in the $ab$ plane. Because of our choice of
azimuth, the incident polarisation was always nearly
\emph{perpendicular} to the $c$ axis.  Consequently, our data were
sensitive only to $\sigma\sigma$-type scattering
\cite{Tensor_formulae} and we can employ the scalar relation in
eq.~(\ref{eq_sf}) with $f'$ and $f''$ being the in-plane
components $f'_{\perp}$ and $f''_{\perp}$ of the anomalous
scattering tensors.

\begin{figure}[tbhp]
\begin{center}
\includegraphics[width=8cm,bbllx=73, bblly=394 ,bburx=341,
bbury=664,angle=0,clip=]{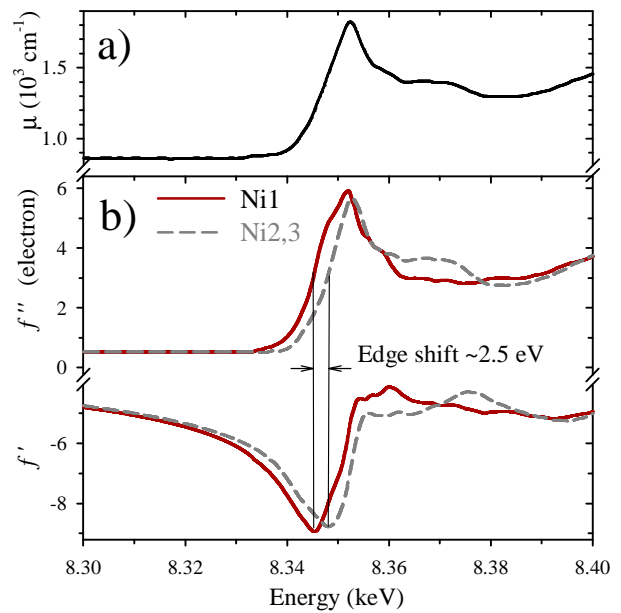} \caption{\label{fig_fprime}
(color online) a) Energy-dependence of the absorption coefficient
$\mu$ obtained from transmission data. b) Energy-dependence of the
empirically-extracted real and imaginary anomalous atomic
scattering factors for Ni1 (solid line) and Ni2,3 (dashed line)
obtained from a best fit to measured spectra as shown in Fig.\
\ref{fig_spectra}.}
\end{center}
\end{figure}

\begin{figure}[tbhp]
\begin{center}
\includegraphics[width=8cm,
bbllx=192,bblly=363,bburx=492,bbury=528,angle=0,clip=] {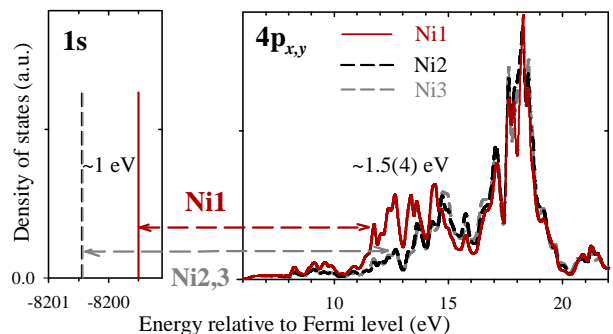}
\end{center}
\vspace{-0.75cm} \caption{(color online) LDA calculation of
density of states (DOS) for the initial and final electron states
in the transition $1s \rightarrow 4p_{x,y}$ (horizontal dashed
arrows) probed by the present resonant Ni K-edge scattering
experiments. Solid/dashed curves show the DOS for
electron-rich/depleted Ni1/Ni2,3 sites. The combined effect is a
predicted transition energy shift of $2.5(4)$ eV higher for Ni2,3
compared to Ni1, in excellent agreement with the experiment.}
\label{fig_dos}
\end{figure}

Remarkably, the complex multiple-peak structure of the spectra in
Fig.~\ref{fig_spectra} can be reproduced quantitatively (in
absolute units) by a model (solid lines) based on an unbiased
reconstruction of the atomic scattering factors for the Ni1 and
Ni2,3 (Fig.\ \ref{fig_fprime}b).  In this analysis we assumed
identical scattering factors for both Ni sites on the honeycomb
$f_{\rm{Ni2}}=f_{\rm{Ni3}}$ as their first-neighbor environments
(Ni-O bond lengths and angles) are identical \cite{ewa}. The two
functions $f_{\rm{Ni1}}''(E)$ and $f_{\rm{Ni3}}''(E)$ were
optimised iteratively to best match the observed spectra, under
the constraints of a given \emph{average}
$f''_{\rm{fu}}=f''_{\rm{Ag}}(E)+2f_{\rm{O}}''(E)+[f_{\rm{Ni1}}''(E)+2f''_{\rm{Ni3}}(E)]/3$
per formula unit, which  is related to the  linear absorption
coefficient (see above) via \cite{Chantler} $\mu(E)=(2 Z h c r_{e}
/V_{\rm uc} E)f_{\rm{fu}}''(E)$ , where $Z=6$ is the number of
formula units per unit cell of volume $V_{\rm uc}$, $hc/E$ is the
x-ray wavelength and $r_{e}$ is the classical electron radius. The
\emph{real} parts of the scattering factors $f'$ in
eq.~(\ref{eq_sf}) were obtained via the Kramers-Kronig (KK)
relation \cite{KKrelations,Chantler} from the imaginary parts
$f''$, such that the \emph{individual} 
anomalous scattering factors obey the KK relation.  Starting
values for the profiles $f_{\rm{Ni1,3}}''(E)$ were obtained by
subtracting the intensities $|F(\bm{Q},E)|^2$ such as in Fig.\
\ref{fig_spectra} for various reflections (with the energy
independent terms $\mp A_{\bm{Q}}f^0_{\rm{O}}/\sqrt{3} + \Delta
f_{\rm Ni}^0$  replaced by their measured values off-resonance) to
give the difference $f_{\rm{Ni1}}'(E)-f_{\rm{Ni3}}'(E)$. After
inserting this back into eq.\ (\ref{eq_sf}) one obtains the
modulus $|f''_{\rm{Ni1}}(E)-f''_{\rm{Ni3}}(E)|$. Combining this
with the constraint for the sum
$f''_{\rm{Ni1}}(E)+2f''_{\rm{Ni3}}(E)$ yields a few possibilities
for starting profiles for the two functions $f''_{\rm{Ni1,3}}(E)$
(this is due to the sign uncertainty), but with only one set
satisfying all the constraints above. The best parameterization of
the data (solid lines in Fig.\ \ref{fig_spectra}) is obtained for
the atomic scattering factors in Fig.~\ref{fig_fprime}b), which
show a clear energy shift of the edge of $+2.5(3)$ eV and some
subtle differences at higher energy\cite{Ni2andNi3}. The sign of
the energy shift is fully consistent with an electron-rich Ni1
site\cite{expNiedge, Eshift1s} and is uniquely determined from the
data. In addition, the absolute scaling of the peak intensities
ensures an accurate and reliable determination of the
\emph{magnitude} of the edge shift.

Edge shifts are often taken as directly proportional to changes in
the formal valence of the ion. For octahedral Ni, the
proportionality constant is $\sim0.66$ electrons/eV
\cite{expNiedge}, so an edge shift of 2.5 eV corresponds to a
disproportionation of $\sim1.65$ electrons, in very good agreement
with the expected CO scenario of Ni1$^{2+}$ and
Ni2,3$^{3.5+}$\cite{ewa}. However,  the initial- and final-states
contributions to the edge shift are typically of similar
magnitudes, while only the former are \emph{directly} related to
charge ordering. To estimate the core-level shifts, we determined
the positions of the $4p$ final states from LDA band-structure
calculations\cite{ewa}. These are completely independent of the
details of the electronic structure near the Fermi energy, and are
therefore reliable and unaffected by assumptions made about
electronic correlations.  The $4p_{x,y}$ (degenerate) densities of
states are shown in Fig.\ \ref{fig_dos}(only these  bands are
probed in the present experiments).  The Ni1 and Ni2,3 $4p$ bands
are shifted by $\sim 1.5$ eV, well short of the $2.5(3)$ eV we
observed experimentally, providing clear evidence that a core
level shift of $\sim$ 1 eV is required to reproduce the data.
Further insight can be gained by extracting directly the positions
of the core $1s$ levels on the Ni1 and Ni2,3 sites from the LDA
calculations (left panel, Fig.\ \ref{fig_dos}), which indeed
differ by  $\sim 1$ eV.  This demonstrates that the LDA
calculations are quantitatively consistent with the resonant
scattering experiment, and provides strong evidence of honeycomb
CO with an amount of charge disproportionation in agreement with
LDA.

In summary, we have reported Ni-K-edge resonant X-ray scattering
measurements on a single-crystal of the triangular-lattice metal
$2H$-AgNiO$_2$ which undergoes a spontaneous ordering transition
that we have previously interpreted in terms of charge
disproportionation and honeycomb charge order. We have observed a
large resonant effect on the superstructure reflections.  The rich
energy-dependent structure can be quantitatively accounted for by
interference scattering from the electron-rich and -depleted Ni
sites. By comparing our data with electronic structure
calculations we determined a core-level shift of $\sim$ 1 eV
between Ni sites, providing clear, direct and quantitative
evidence of charge ordering.

We acknowledge support from EPSRC UK and a studentship from the
University of Bristol and Potter foundation (GLP). We thank E.\
Wawrzy\'{n}ska and  M.\ Brunelli for characterising the
polycrystalline sample on Beamline ID31 at the ESRF.




\begin{thebibliography}{99}
\bibitem{correction} Raw intensities were corrected for
Debye-Waller factor (using atomic displacements from
\cite{Soergel05}), Lorenz factor, and absorption using a numerical
beam-path integration routine for a disk-shaped sample with
effective dimensions refined against the measured (off-resonance)
intensity ratios of several main ($hkl$) reflections with $h-k=3n$
and $l$ odd where Ni contributions to $F$ cancel out. The linear
absorption coefficient $\mu(E)$ in Fig.\ \ref{fig_fprime}a) was
extracted from the measured transmission spectrum through a thin
layer of powder distributed on kapton tape. Off-resonance the
analytic approximation for $f'(E)$ and $f''(E)$ was used
\cite{Chantler}.
\bibitem{Tensor_formulae}
For the case of no analyser (as in the present experiments)
eq.~(\ref{eq_sf}) becomes the sum (we drop the $E$ and $Q$
dependence for brevity) $|F|^2=|F|^2_{\sigma\sigma}
+|F|^2_{\sigma\pi}$, where $|F|^2_{\sigma\sigma}=3\left[\mp
A_{\bm{Q}}f^0_{\rm{O}}/\sqrt{3}+\Delta f^0+\Delta
f'_{\perp}+\gamma^2\left(\Delta f'_{\parallel}-\Delta
f'_{\perp}\right)\right]^2 + 3\left[\Delta
f_{\perp}''+\gamma^2\left(\Delta f''_{\parallel}-\Delta
f''_{\perp}\right)\right]^2$ and $|F|^2_{\sigma\pi}=3
(\gamma\gamma')^2 \left[\left(\Delta f_{\parallel}' -\Delta
f'_{\perp}\right)^2+\left(\Delta f_{\parallel}'' -\Delta
f''_{\perp}\right)^2\right]$ where the difference terms are
$\Delta f''_{\perp}
=f''_{\perp}({\rm{Ni1}})-f''_{\perp}({\rm{Ni3}})$ and so on. Here
$\gamma$, $\gamma'$ are projections onto the crystal
$\bm{c}^*$-axis of the incident x-ray beam polarization vector
$\bm{\sigma}$ (which is perpendicular to the scattering plane) and
of the unit vector $\bm{\pi}$ (within the scattering plane and
$\perp$ to $\bm{k_f}$).
\bibitem{Ni2andNi3}
The assumption $f_{\rm{Ni2}}=f_{\rm{Ni3}}$ was further justified
by the observed agreement with the data at supercell peaks with
$l$ even, (106), where both Ni2 and Ni3 contribute.
\end{thebibliography}
\end{document}